\documentclass[aps,pre,amsmath,amssymb,reprint,numeric]{revtex4-1}
\usepackage[english]{babel}
\usepackage{float}
\usepackage{amsmath, amssymb, amsthm, amsfonts}
\usepackage[applemac]{inputenc}
\usepackage[pdftex]{graphicx} 
\usepackage{color}
\usepackage{bm}
\usepackage[normalem]{ulem}
\usepackage{multirow}

\begin{document}
\title{On the scaling of avalanche shape and activity power spectrum in neuronal networks}

\author{Manoj Kumar Nandi$^1$, Alessandro Sarracino$^1$, Hans J. Herrmann$^{2,3}$, Lucilla de Arcangelis$^1$}
\email{lucilla.dearcangelis@unicampania.it}

\affiliation{$^1$ Department of Engineering, University of Campania "Luigi Vanvitelli"‚ 81031 Aversa (Caserta), Italy\\
$^2$ PMMH, ESPCI, 7 quai St. Bernard, Paris 75005, France\\
$^3$ Departamento de Fisica, Universidade Federal do Cear\'{a}, 60451-970, Fortaleza, Cear\'{a}, Brazil}

\begin{abstract}
Many systems in Nature exhibit avalanche dynamics with scale-free features. A general scaling theory has been proposed for critical avalanche profiles in crackling noise, predicting the collapse onto a universal avalanche shape, as well as the scaling behaviour of the activity power spectrum as Brown noise. Recently, much attention has been given to the profile of neuronal avalanches, measured in neuronal systems in vitro and in vivo. Although a universal profile was evidenced, confirming the validity of the general scaling theory, the parallel study of the power spectrum scaling under the same conditions was not performed. The puzzling observation is that in the majority of healthy neuronal systems the power spectrum exhibits a behaviour close to $1/f$, rather than Brown, noise. Here we perform a numerical study of the scaling behaviour of avalanche shape and power spectrum for a model of integrate and fire neurons with a short-term plasticity parameter able to tune the system to criticality. We confirm that, at criticality, the average avalanche size and the avalanche profile fulfil the general avalanche scaling theory. However, the power spectrum consistently exhibits Brown noise behaviour, for both fully excitatory networks and systems with 30\% inhibitory networks. Conversely, a behavior closer to $1/f$ noise is observed in systems slightly off-criticality. Results suggest that the power spectrum is a good indicator to determine how close neuronal activity is to criticality.
\end{abstract}

\maketitle

\section{introduction}

Beside the well-known rich phenomenology, i.e., instabilities and metastability
transitions~\cite{freeman2006dynamics,freeman2005metastability,deco2012ongoing},
synchronization~\cite{nunez2001spatial,benar2019differences}, the
presence of multiple spatio-temporal scales~\cite{deco2011emerging,robinson2003neurophysical} and long-range temporal correlations~\cite{Linken01}, spontaneous brain activity exhibits bursts of activity, named neuronal avalanches. These have been first detected
{\it in vitro} in organotypic cultures from coronal slices of rat cortex \cite{beggs2003} and in dissociated neurons from rat hippocampus and cortex \cite{mazzoni,pasquale}  or leech ganglia \cite{mazzoni}. Next, neuronal avalanches
have been identified {\it in vivo} in rat cortical layers
during early post-natal development \cite{gireesh2008neuronal}, in the cortex of awake adult rhesus monkeys \cite{petermann2009spontaneous}, as well as in the resting state of the human brain by means of non-invasive techniques such as magneto-encephalography (MEG) \cite{shriki2013}. The statistical properties of neuronal avalanches have been object of intensive investigation both experimentally and numerically \cite{prl06,levina,Laurence2018}, focusing mainly on the scaling behaviour of the avalanche size and duration distributions, $P(S)$ and $P(T)$. Consistent evidence indicates a scaling behaviour in the universality class of the mean field branching model \cite{zapperi1995,kessenich16}, namely $P(S)\propto S^{-\alpha}$ with $\alpha\simeq1.5$ and $P(T)\propto T^{-\tau}$ with $\tau \simeq 2.0$, even if the existence of a different universality class has been also proposed in the literature \cite{fonte}.

Avalanching is a widespread phenomenon, occurring in systems where many degrees of freedom interact under slow drive, which had, as first prototype, the model of Gutenberg-Richter \cite{gutric} for earthquake occurrence in the 50s and became a general paradigm after the seminal work of Per Bak \cite{bak,jensen}. Within this context, a real breakthrough was the formulation of a general scaling theory for avalanche phenomena at the critical point, encompassing the scaling behaviour of most relevant properties in the process. This scaling theory, initially formulated for the Barkhausen noise \cite{sethna2001,kuntz}, has turned out to be extremely general and found in a variety of different phenomena, from plastic deformations \cite{laurson} to earthquakes \cite{mehta}. Among a number of scaling relations for different quantities \cite{kuntz}, some scaling laws have been considered in the literature as an indicator to determine if a system acts at the critical point. In particular, the scaling of the average avalanche size $<S>$ versus its duration $T$ \cite{kuntz,sethna2001,papanikolaou}, $<S>\sim T^{\gamma}$, with  the exponent $\gamma$ related to the exponents of the avalanche distributions
\begin{equation}
\gamma=\frac{\tau-1}{\alpha-1}.
\label{sethna}
\end{equation}
This exponent $\gamma$ is also predicted to control the collapse of the profile of avalanches with different durations, as well as the scaling of the signal Power Spectral Density (PSD), $S(f)\sim f^{-\gamma}$ \cite{kuntz}. Moreover, other features of the avalanche profile have received a wide interest. In particular, the avalanche shape, not necessarily symmetric in the scaling theory \cite{sethna2001}, has been found to depend on the interaction kernel, namely asymmetry appears when the interaction is not fully non-local, reflecting broken time-reversal symmetry in the avalanche dynamics \cite{alava}. Symmetric profiles are found in Barkhausen noise \cite{sethna2001,kuntz} and in mean-field systems \cite{papanikolaou}, however considering inertial effects leads to leftward asymmetry (positive skewness) \cite{zapperi2007}, whereas rightward asymmetry (negative skewness) is observed for increasing interaction range \cite{alava}. Interestingly, the symmetry of avalanche shape in experimental neuronal avalanches shows a variety of features: From an almost symmetric shape in cortex slices \cite{beggs12} and in non-human primates \cite{miller}, provided that the modulation of $\gamma$-oscillations is carefully taken into account, to a leftward asymmetric shape in zebrafish larvae \cite{ponce}. This confirms that the symmetry of the shape is not a necessary requirement for criticality, as already observed in \cite{sethna2001}.

Interestingly, experimental and numerical studies aiming at verifying the validity of the avalanche scaling theory for neuronal avalanches, mainly focused on the scaling of the average size with duration and the collapse of avalanche profiles onto a universal curve \cite{beggs12,miller,ponce,plos}. Little attention has been given to the parallel investigation of the scaling behaviour of the activity power spectrum, which is predicted by \cite{kuntz} to exhibit Brown noise behaviour, i.e., $S(f)\sim f^{-\beta}$, with $\beta=\gamma=2$ (Eq.~(\ref{sethna})). A variety of experimental studies on different signals, as EEG, MEG, resting state fMRI, as well as the LFPs of spontaneous cortical activity, have evidenced the presence of effective power law regimes in the Power Spectral Density (PSD) \cite{novikov,bedard,dehghani,prichard,zarahn}, which are the background for peaks at characteristic frequencies corresponding to different brain modes. In the majority of studies, a behaviour closer to $1/f$, rather than Brown noise, has been detected. For instance,
the PSD scaling behaviour in the human eyes-closed and eyes-open resting
EEG \cite{prichard} provides at low frequencies (0.5–8 Hz) $\beta\sim 1.32$ for the eyes-closed condition and $\beta\sim 1.27$ in the eyes-open condition, with the exponent varying across
brain regions with a standard error of 20\%. A similar analysis of both EEG and MEG signals has recently proposed \cite{dehghani} an average
slope $\beta=1.06\pm 0.29$, varying in different brain areas. Conversely, the exponent value $\beta\sim 2$ is typically found in epileptic patients, in the range [2.2, 2.44] in the awake state and [1.6, 2.87] in the Slow Wave Sleep \cite{he}. In addition, a numerical study of a self-organized model for neuronal activity has evidenced that the exponent $\beta$ depends on the fraction of inhibitory neurons, crossing over from Brown noise for fully excitatory networks to $1/f$ behaviour for 30\% inhibitory neurons \cite{chaos}, typical of mammal brains.

Finally, it is important to remark that there is evidence in the literature \cite{TD17,TD21} that the observed scaling behavior for avalanches can be also  obtained without invoking the criticality hypothesis. Recent investigation has been triggered by the results in \cite{fonte} showing that data,
ranging from freely moving to anesthetized mammals, surprisingly show scaling behavior in a specific activity regime, extending original results which focused on spontaneous brain activity. It has therefore been suggested that power laws can emerge even in models which are not critical \cite{TD21} and do not satisfy Eq.~(\ref{sethna}). Therefore, the criteria implemented in \cite{fonte} to confirm criticality would be not sufficient to discriminate critical from non-critical systems. Indeed, the emergence of power law distributions has been recently related to the presence of fluctuating variables in the process, such as input stimuli leading to neuronal firings at different rates \cite{schwab}. 

Here we address the issue of scaling for neuronal avalanches within the context of an integrate and fire neuronal network model including short and long-term plasticity \cite{Laurence2018}. The model presents the advantage of being able to tune the system at and far from criticality, allowing the investigation of the scaling behaviour in a wider region of phase space. In particular, we will study the scaling properties of the avalanche shape in parallel with the activity PSD, to identify the conditions for the validity of the avalanche scaling theory. The analysis is performed for different percentages of inhibitory neurons, unveiling their crucial role in the PSD scaling.

\section{Network model}
We consider a neuronal network consisting of $N$ neurons, randomly placed in $3d$
and connected by directed synapses, with a fraction $P_{in}$ of inhibitory neurons. The outgoing connections are assigned according to the scale free degree distribution $P(k_{out})\propto k_{out}^{-2}$ found experimentally for functional networks \cite{scale_free}, with $k_{out}\in [2,100]$. The connections between two neurons are established using a distance-dependent probability, $P(r)\propto e^{-r/r_0}$, where $r$ is the Euclidean distance between two neurons and $r_0=5$ a characteristic connectivity range \cite{Roerig2002}. We assign to each synaptic connection between neuron $i$ and $j$ an initial random strength $g_{ij}$ uniformly chosen in the interval $[0.4,0.6]$. A recent study \cite{Bonifazi2009} has shown that in scale-free functional networks the inhibitory neurons are typically hubs, we therefore choose the inhibitory neurons among the highly connected ones, $k_{out}>5$. 
We implement in the model two plasticity mechanisms, the short- and long-term plasticity. The first one models the dynamics of synaptic resources that are used in all firing events and need to be restored in the presynaptic terminal. By tuning the efficiency in restoring neurotransmitter resources, the model tunes the network at the critical point. This plasticity mechanism is always active during the avalanche dynamics. Conversely, the long-term plasticity is a Hebbian mechanism that models the experience of the neuronal systems by sculpting the synaptic strengths according to their activity. This plasticity mechanism mimics the age of the neuronal system and modifies the synaptic connections starting from an initial random configuration. For this reason, the long-term plasticity is active only during an initial training period, during which avalanche measurements are not monitored, in order to store into the synaptic strength information about previous activity. The implementation of both plasticity measurements, acting on very different time scales (ms the first, up to years the second) is motivated by the need to include in the model fundamental experimental evidence for neuronal dynamics.
More precisely, each neuron $i$ is characterized by a membrane potential $v_i$, whose initial values are distributed uniformly in the interval $[0.5,1.0]$.
At each time step all neurons with a potential exceeding a
threshold value $v_i \geq v_c = 1.0$ fire, leading to the release of a fraction of neurotransmitter $\delta u$ at all synapses and a change in the potential of the connected postsynaptic neurons $j$ according to the following equations \cite{Laurence2018,Raimo2020RoleOI}

\begin{gather*}
 	v_j(t+1)=v_j(t)\pm v_i(t)u_i(t)\delta u g_{ij}\\
 	u_i(t+1)=u_i(t)(1-\delta u)\\
 	v_i(t+1)=0
 	\label{dynamics}
\end{gather*}

where the $\pm$ sign stands for excitatory or inhibitory presynaptic neuron and $u_i$ is the amount of releasable neurotransmitter at neuron $i$, which is initially set equal to one for all neurons. After a neuron fires, it enters into a refractory state for $t_r=1$ time step during which it is unable to elicit any signal. The activity propagates and stops when the potential of all neurons is below $v_c$. The present model is the discrete time version of the integrate and fire model by ref. \cite{levina}. Here the unit time step is of the order of 10ms and represents the time elapsed between the elicitation of the action potential and the change in membrane potential at the postsynaptic neuron.
The amount of available neurotransmitter at each neuron $u_i$ is called readily-releasable pool \cite{rosenmund,kaeser} and  only a 5\% fraction is released at each firing \cite{rizzoli,ikeda} ($\delta u=0.05$), according to short-term synaptic plasticity.
When all neurons are below firing threshold, activity stops and can be triggered again by setting any random neuron to its threshold value, generating another burst of firing neurons, called neuronal avalanche. We measure for each avalanche the size $S$ as the number of firing neurons, independently of their firing rate, and the duration $T$ as the number of time steps in the activity propagation. During an avalanche, because of short-term plasticity, the available neurotransmitter decreases constantly. Therefore to sustain further activity a recovery mechanism is needed.  Since the synaptic recovery takes a time of the order of seconds \cite{markram,abrahamsson,armbruster2011}, whereas synaptic transmission acts on the scale of milliseconds \cite{beggs2003,lombardi}, the available neurotransmitter recovery is implemented at the end of each avalanche propagation as $u_i=u_i+\delta u_{rec}$ for all neurons. Here $\delta u_{rec}$ is a tunable parameter which determines whether the network is in a subcritical, critical or a supercritical state, i.e., in the critical state the cutoff in the scale free avalanche size distribution correctly scales with the system size $N$ \cite{Laurence2018,Raimo2020RoleOI}, showing that the power-law is due to criticality.

The synaptic network structure is sculpted by the activity-dependent long-term plasticity, following the principles of Hebbian plasticity \cite{hebb}. Whenever a neuron $i$ fires, all synaptic strengths $g_{ij}$ to post-synaptic neurons $j$ are strengthened proportionally to the potential variation induced in the post-synaptic neuron, as $g_{ij}(t+1)=g_{ij}(t)+\delta g_{ij}$, where
 $\delta g_{ij}=\varepsilon |(v_j(t+1)-v_j(t))|$ and
$\varepsilon=0.04$ is a parameter controlling the strength of plastic
adaptation. 
Conversely, at the end of each avalanche, all $g_{ij}$ are reduced by the
average increase in synaptic strength per bond, $g_{ij}=g_{ij}-\sum \delta g_{ij}/N_s$, where $N_s$ is the total number of synapses in the network. 
Synaptic connections whose strength becomes smaller than $g_{min}=10^{-5}$ are pruned, i.e., permanently removed from the network.
In order to modulate the initially random strengths, the long-term plastic adaptation is either implemented for a fixed number of stimulations $N_p=10000$  or stopped at the first synaptic pruning. 
We have studied systems with $N=16000$ neurons in a cube of side $L=100$ and data are averaged over 5000 network configurations.

 \begin{figure}
	\includegraphics[width=0.5\textwidth]{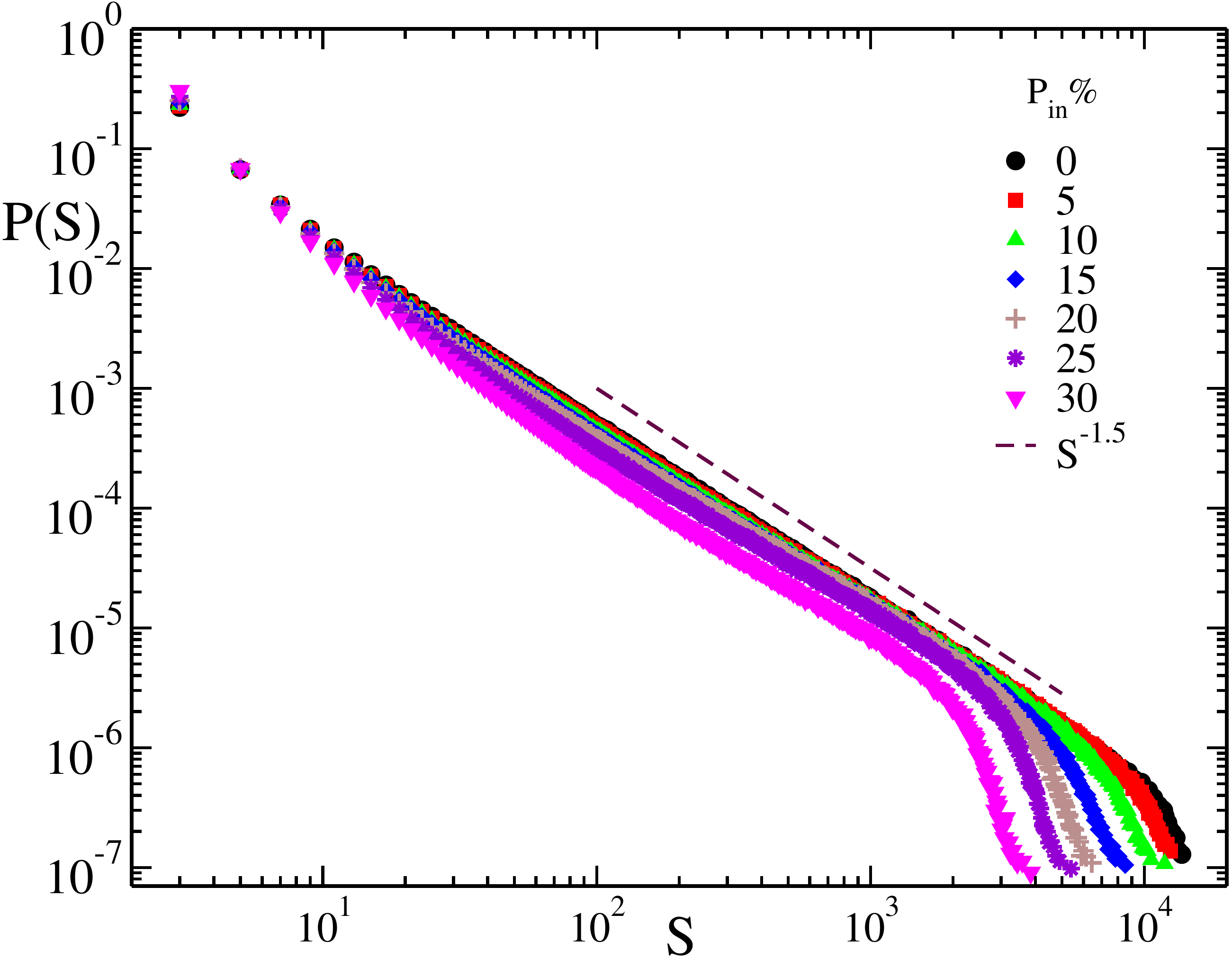}
	\includegraphics[width=0.5\textwidth]{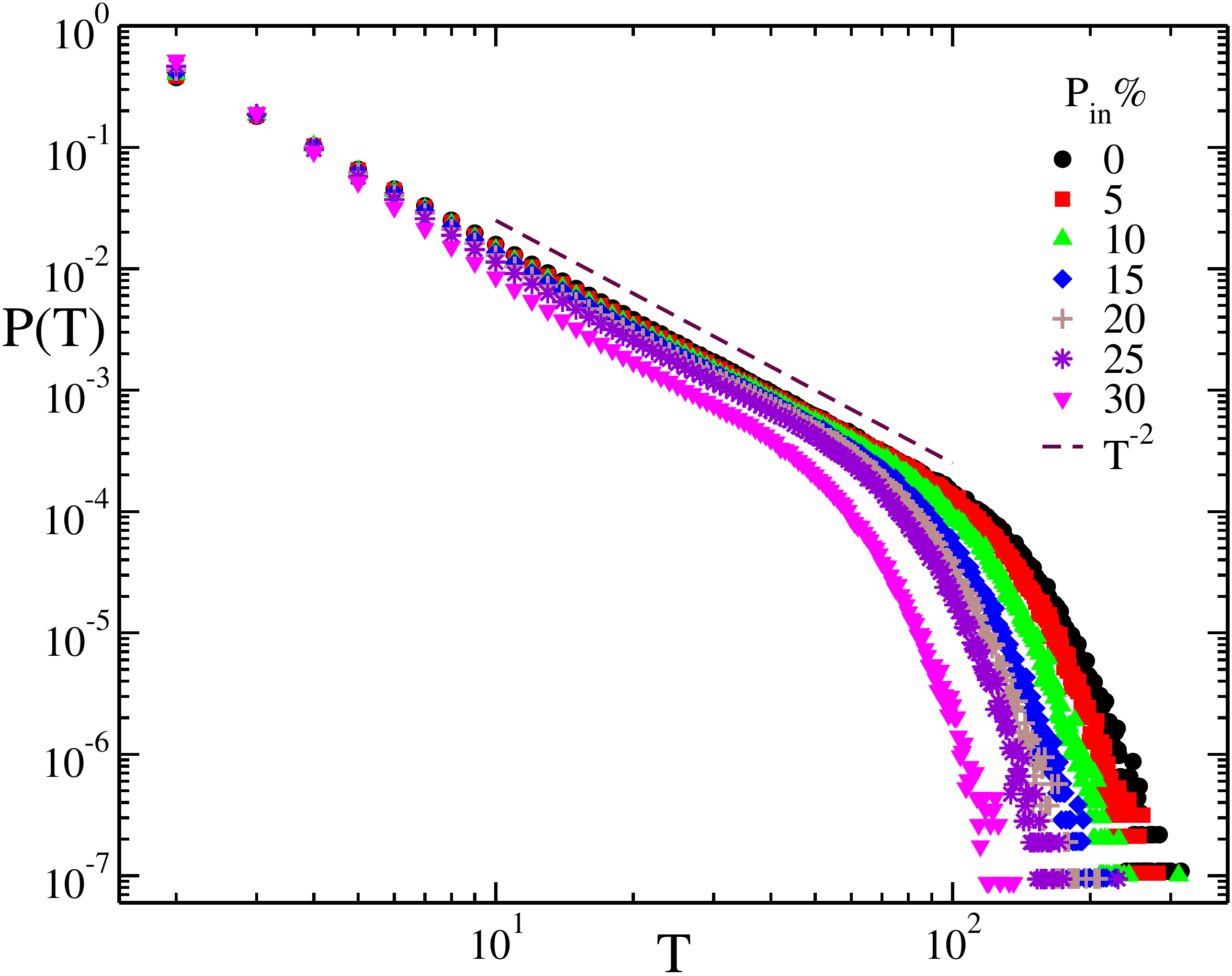}
	\caption{Avalanche size (top panel) and duration (bottom panel) distributions for 5000 configurations of a network of $N = 16000$ neurons for $P_{in} = 0, 5, 10, 15, 20, 25$ and 30\% in the critical state.}
	\label{dis_critical}
\end{figure}

\section{Distribution of avalanche sizes and durations}

As mentioned in the description of the model, it is possible to tune the system in different activity states, subcritical, critical and supercritical, by adjusting the parameter $\delta u_{rec}$ \cite{Laurence2018,Raimo2020RoleOI} controlling the efficiency in neurotransmitter recovery. Since inhibitory neurons hamper the activity propagation and limit large avalanches, by increasing their fraction the system will move away from the critical state into the subcritical region. 
Here we consider networks in the critical state, for which  $\delta u_{rec}$ is appropriately increased for increasing fractions of inhibitory neurons. 
More precisely, we slowly increase the value of $\delta u_{rec}$ moving from a subcritical regime (exponential size distribution) to a regime where a power law behavior is detected in the size distribution. The critical value of $\delta u_{rec}$ is identified as the largest value providing a power law with the cutoff as close as possible to $N$. We also verify that no bump appears in the tail of the distribution, sign of an excess of large avalanches.
In previous studies it has been verified that activity can be tuned to be genuinely critical, i.e., the cutoff in the scale free avalanche size distribution correctly scales with the system size $N$ \cite{Laurence2018,Raimo2020RoleOI}.
The distribution of avalanche sizes $P(S)$ and durations $P(T)$  are shown in  Fig.\ref{dis_critical} for different percentages of inhibitory neurons, each time tuning $\delta u_{rec}$ to the critical state. 
The size distribution exhibits a power-law behaviour $P(S)\propto S^{-\alpha}$ with $\alpha\simeq1.5$, quite independently of the percentage of inhibitory neurons $P_{in}$, followed by an exponential cutoff which moves towards smaller sizes $S$ as we increase the percentage of inhibitory neurons, hindering the occurrence of large avalanches. 
The distribution of durations also shows a power-law behaviour $P(T)\propto T^{-\tau}$ with $\tau \simeq 2.0$ followed by an exponential cutoff at large avalanche durations depending on $P_{in}$. Both scaling exponents are in agreement with experimental values \cite{beggs2003,petermann2009spontaneous,fransson2012early,shriki2013} and consistent with the mean field branching model universality class \cite{zapperi1995} (see Table 1).

\begin{figure}
	\includegraphics[width=0.5\textwidth]{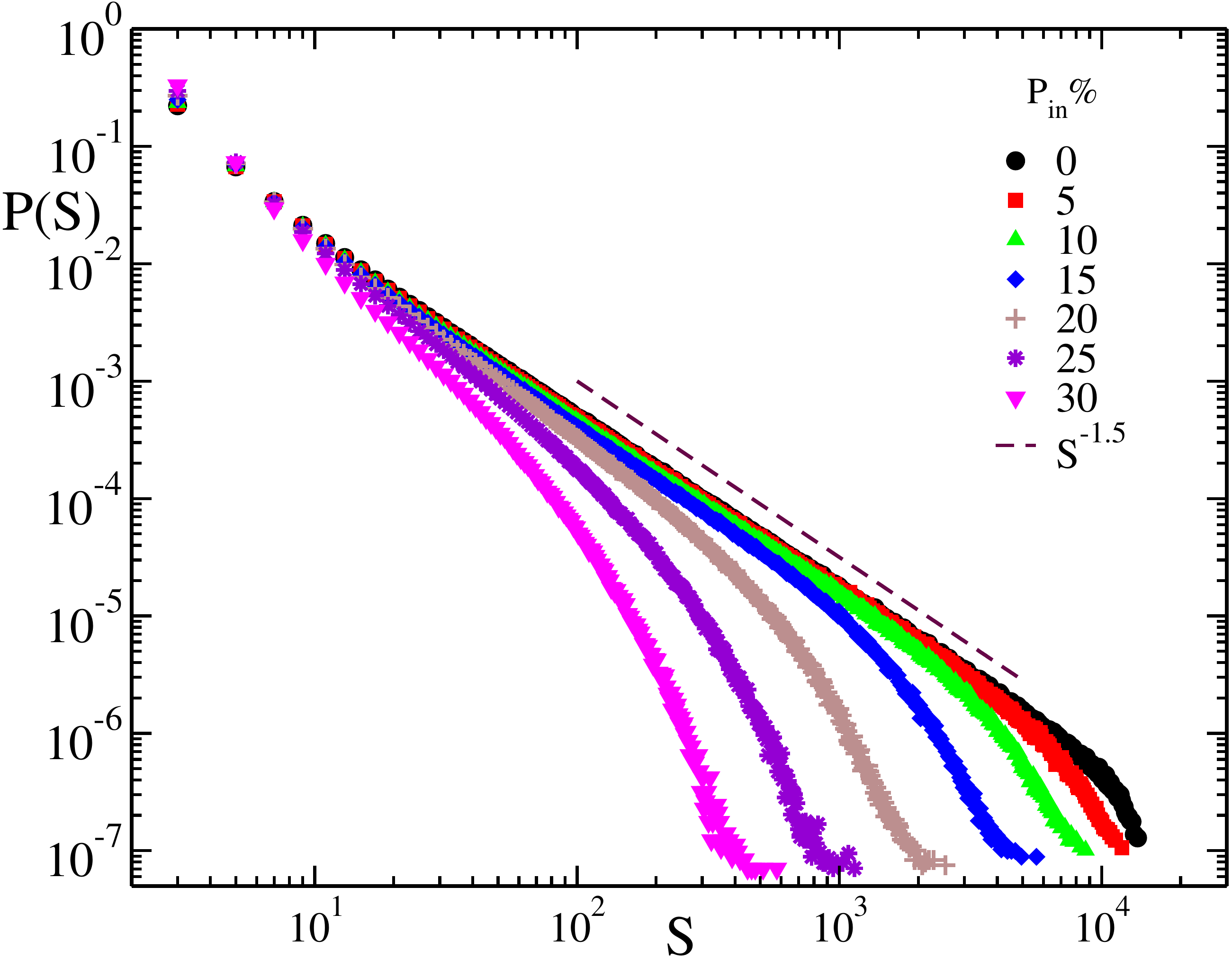}
	\includegraphics[width=0.5\textwidth]{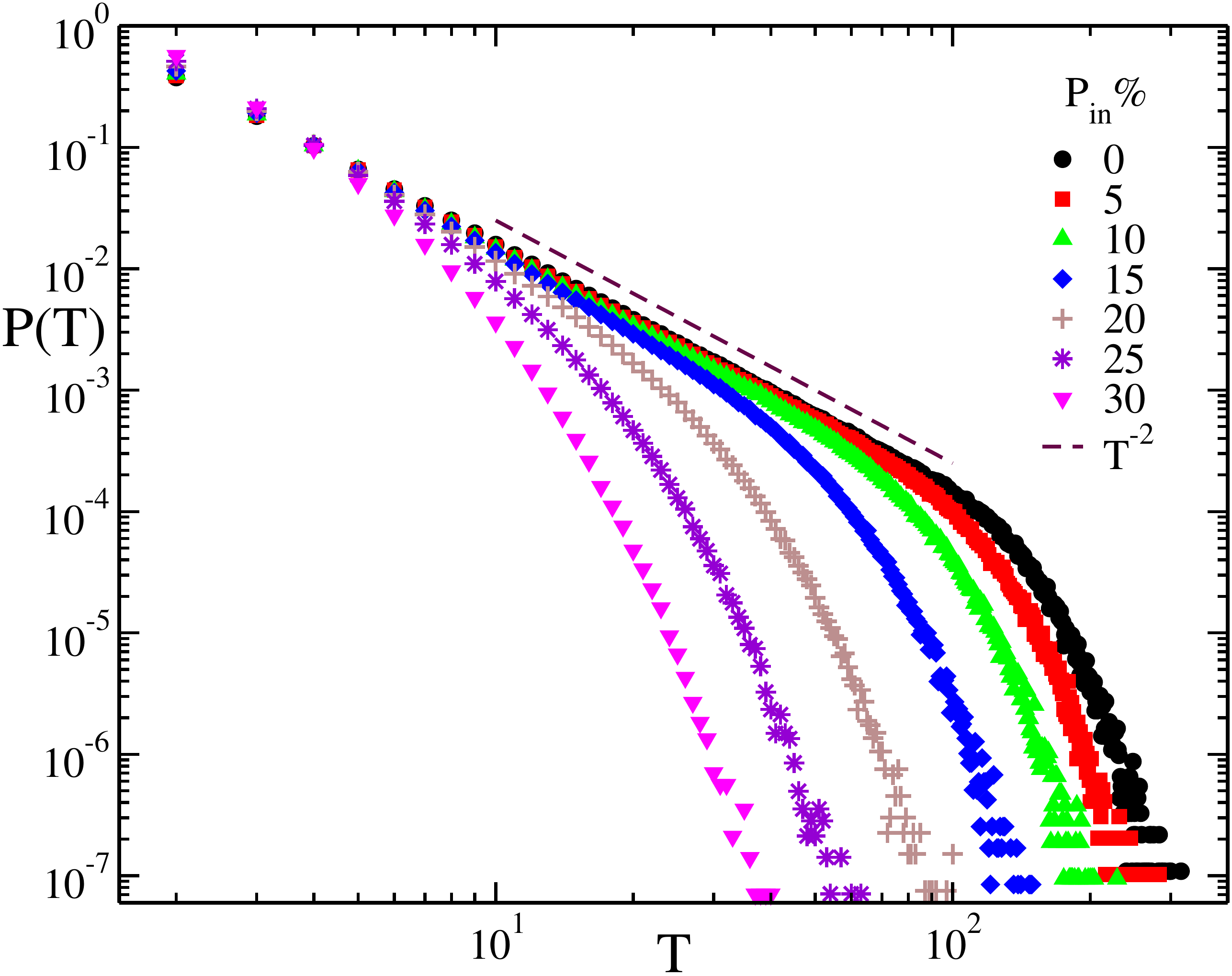}
	\caption{Avalanche size (top panel) and duration (bottom panel) distributions for 5000 configurations of a network of $N = 16000$ neurons for $P_{in} = 0, 5, 10, 15, 20, 25$ and 30\% for fixed $\delta u_{rec}=0.001$, value corresponding to the critical state for $P_{in}=0\%$.}
	\label{dis_subcritical}
\end{figure}

The critical exponents for the avalanche distributions have been also found in a self-organized neuronal network model without short term plasticity, namely in absence of any tuning parameter \cite{prl06,chaos}. For this model with different $P_{in}$ the size distribution has been found to follow the scaling form $P(S)=S^{-\alpha}f(S/P_{in}^{-\theta})$, where $\alpha=1.5$ and the cutoff size scales with $P_{in}$ with an exponent $\theta\simeq 2.2$ for scale free networks. In self-organized models the fraction of inhibitory neurons strongly affects the extension of the scaling regime, however data collapse confirms the value of $\alpha$ by appropriately considering the cutoff scaling. 
To understand the role of inhibitory neurons in the present model, which is not self-organized but tuned at criticality, we first determine the value of $\delta u_{rec}$ which sets a purely excitatory system ($P_{in}=0\%$) in the critical state. Then we fix the value of $\delta u_{rec}$ and increase the percentage of inhibitory neurons in the system. By doing so, the system moves away from the critical point, into the subcritical regime, which leads to a decrease of the scaling regime in the distributions with the exponential cutoff moving towards smaller avalanche sizes (Fig.\ref{dis_subcritical} top panel). Similar behaviour is also observed for the distribution of avalanche durations (Fig.\ref{dis_subcritical} bottom panel, see Table 1).
\begin{figure}
	\includegraphics[width=0.5\textwidth]{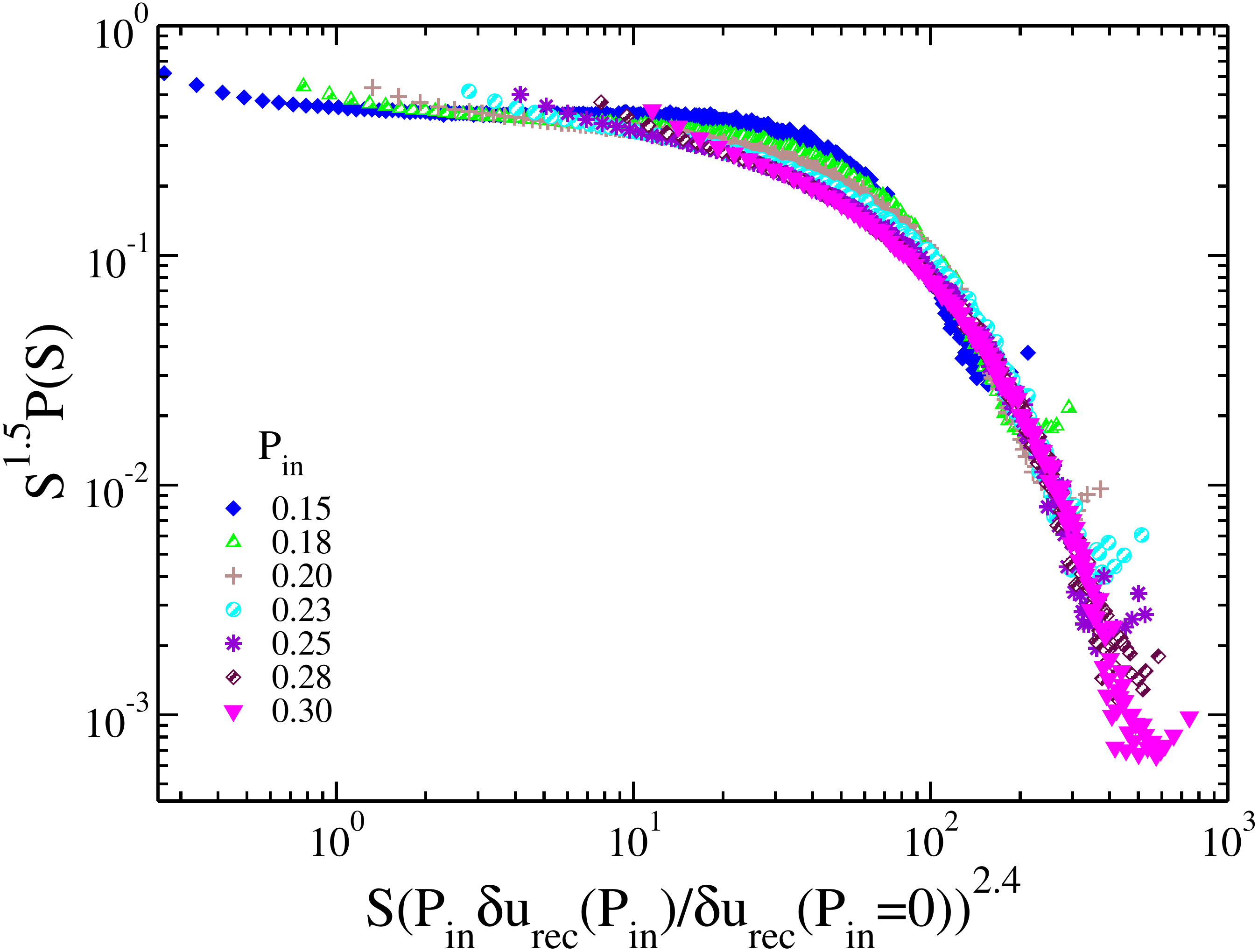}
	\caption{Collapse of avalanche size distributions off criticality onto a universal curve by plotting $S^{\alpha}P(S)$ vs $S(P_{in}\delta u_{rec}(P_{in})/\delta u_{rec}(P_{in}=0))^{\theta}$, with $\alpha=1.5$ and $\theta=2.4$.}
	\label{data_collapse}
\end{figure}

\begin{figure}
	\includegraphics[width=0.5\textwidth]{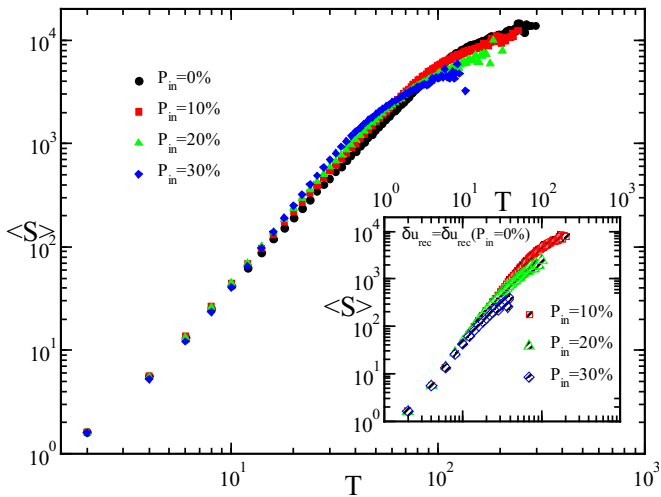}
	\caption{Average avalanche size $<S>$ vs. duration T. The slope for all values of $P_{in}$ is close to 2. Inset: The same quantity for systems not tuned at criticality, i.e. for fixed $\delta u_{rec}=0.001$ and $P_{in}=10\%,20\%,30\%$ }
	\label{avgS_T}
\end{figure}

\begin{figure*}[t]
	\centering
	\includegraphics[width=0.32\textwidth]{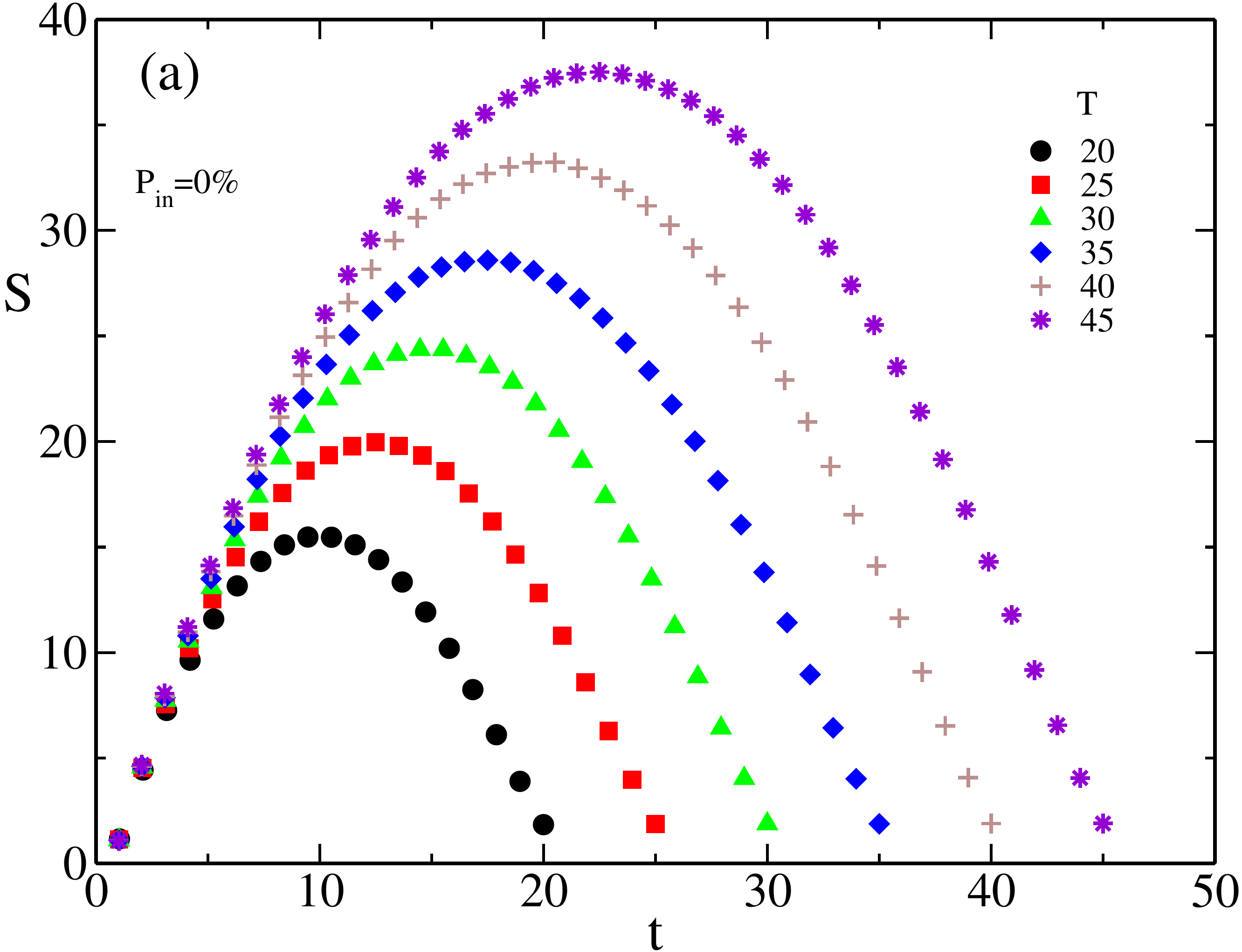}
	\includegraphics[width=0.32\textwidth]{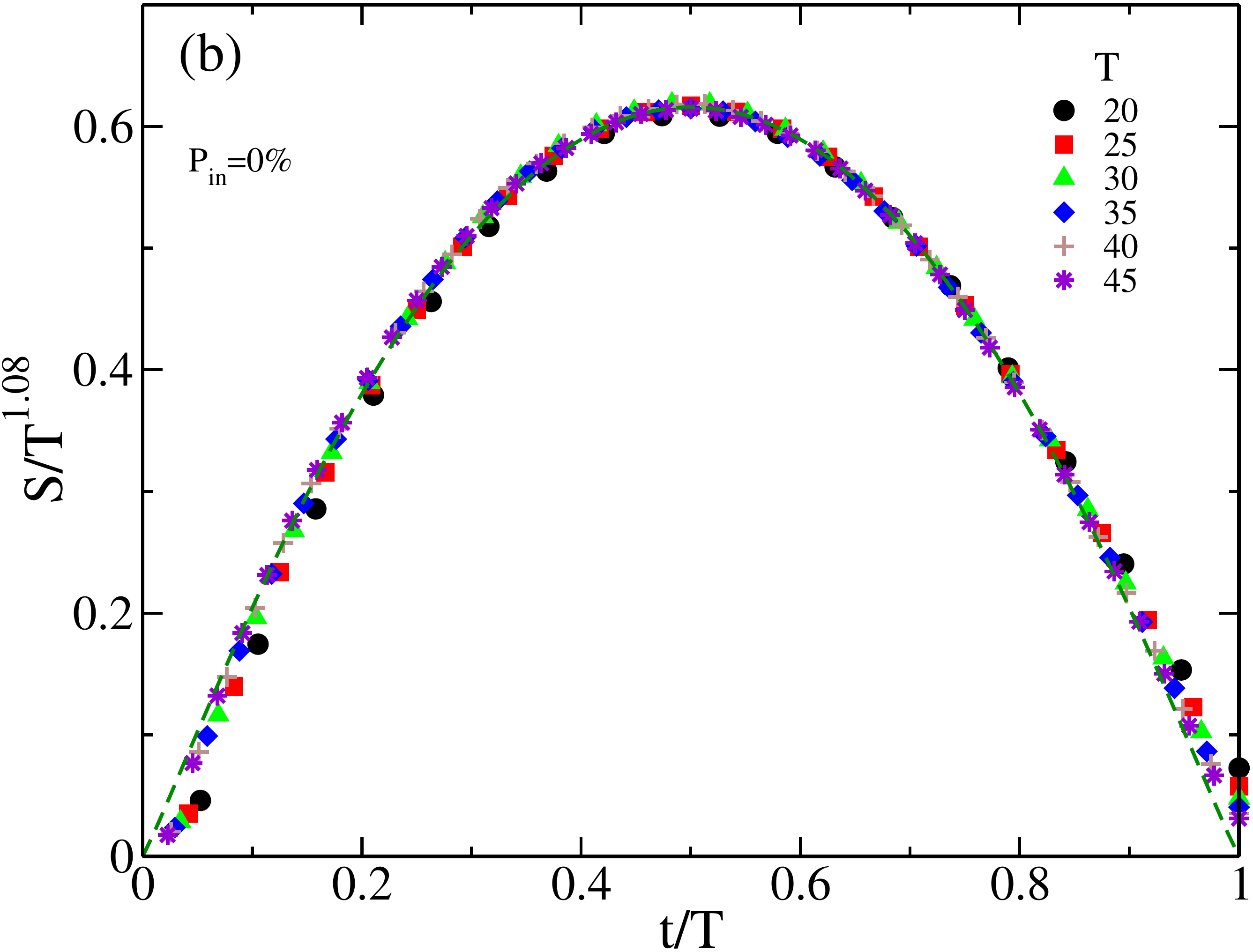}
	\includegraphics[width=0.32\textwidth]{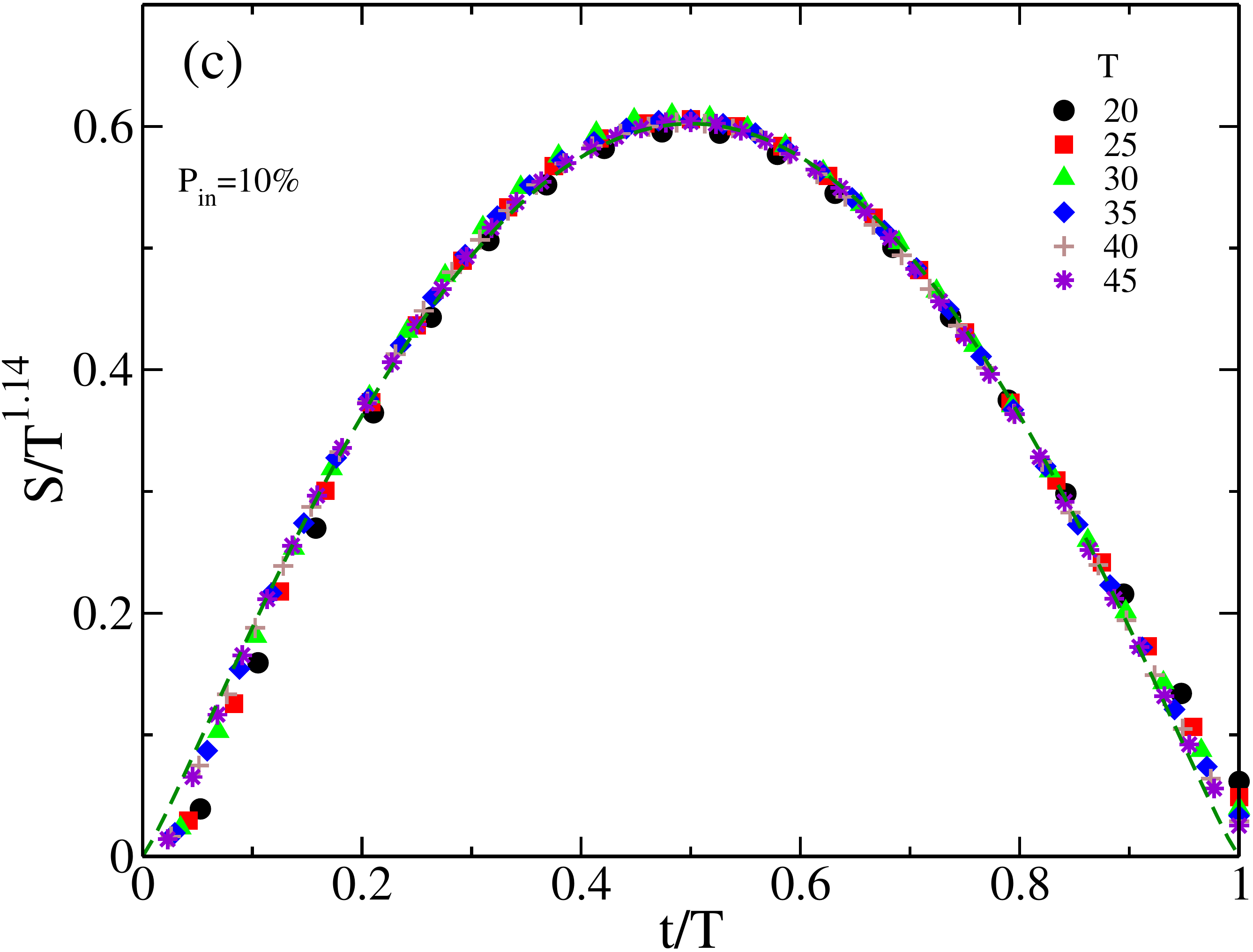}	
	\includegraphics[width=0.32\textwidth]{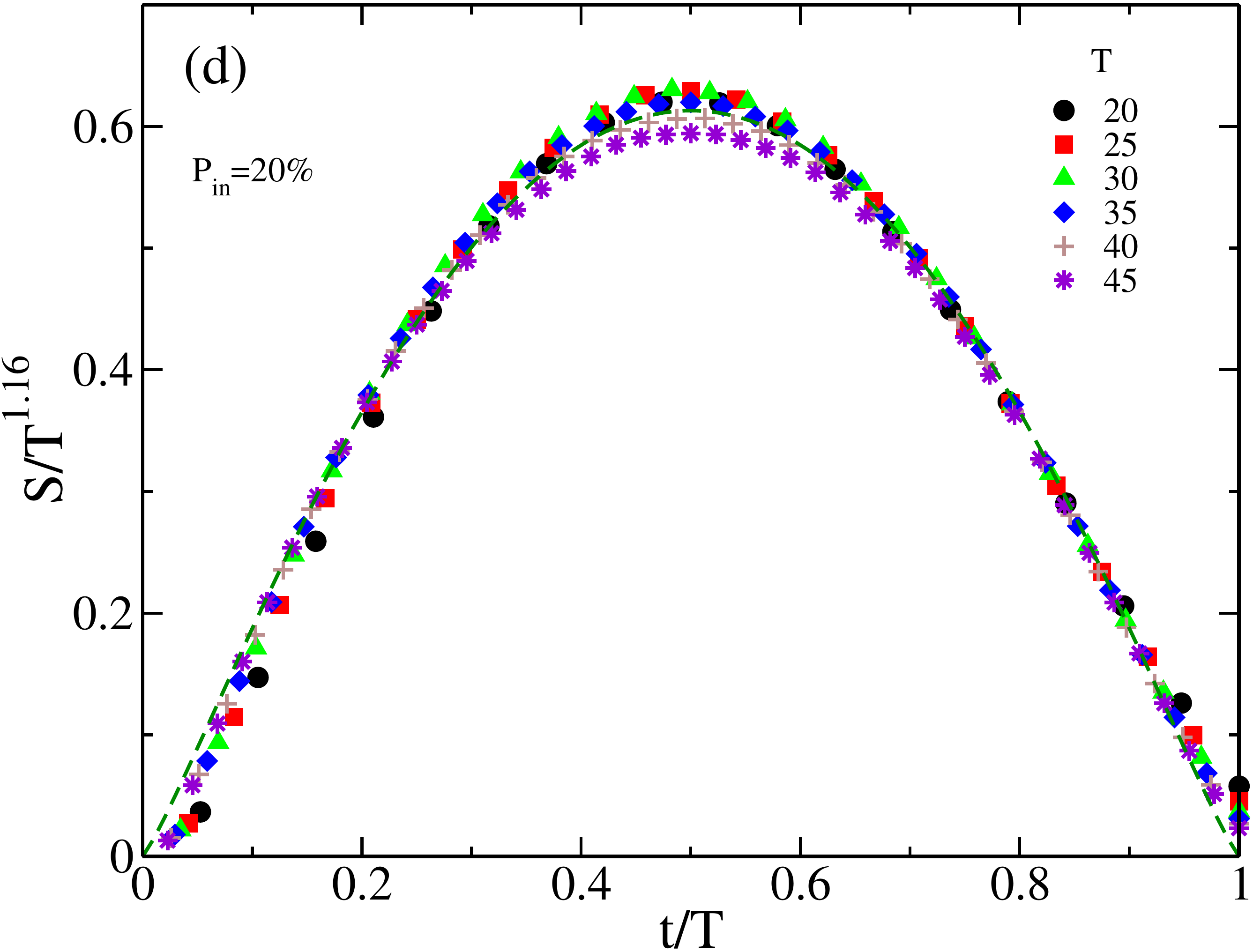}	
	\includegraphics[width=0.32\textwidth]{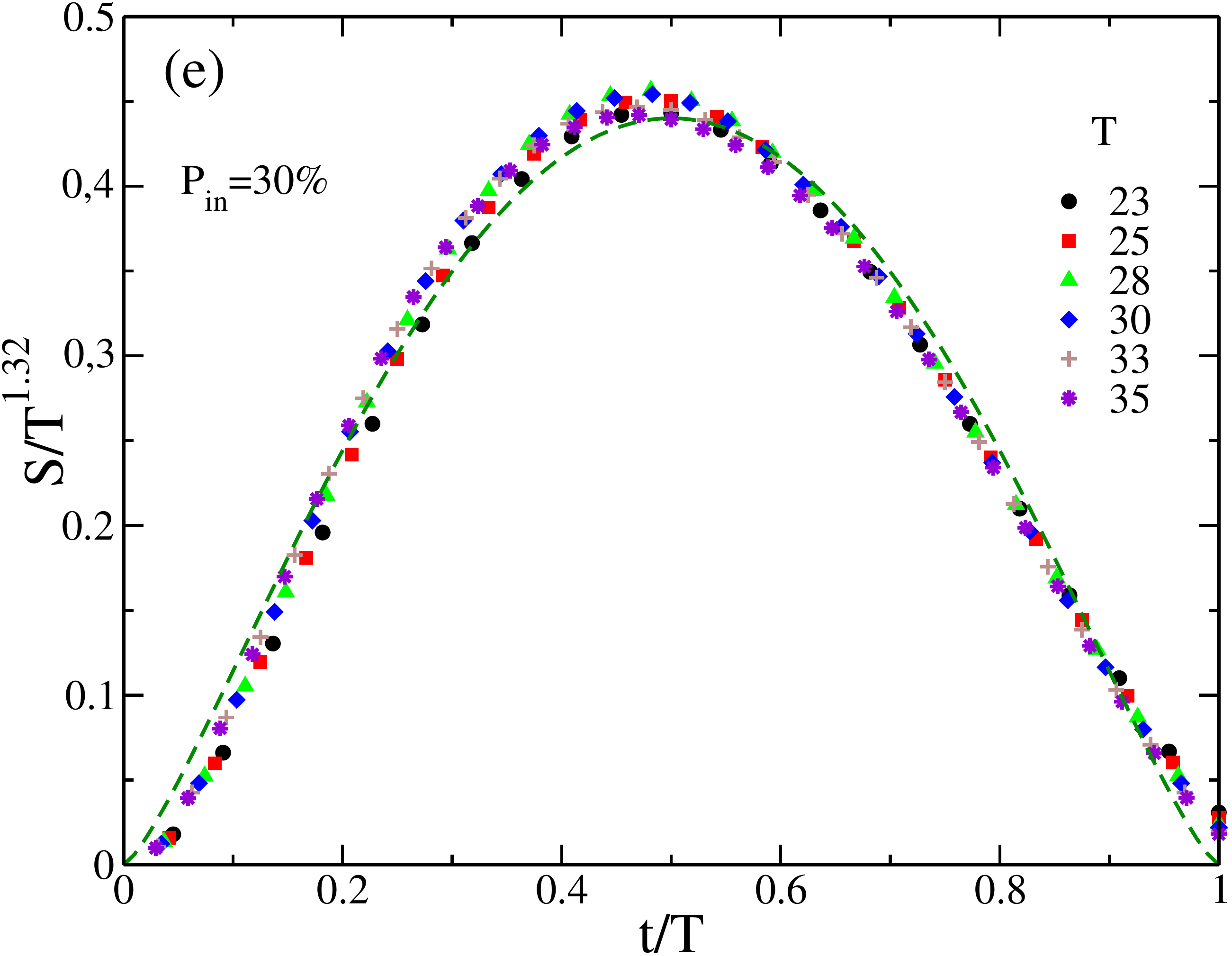}
	\includegraphics[width=0.32\textwidth]{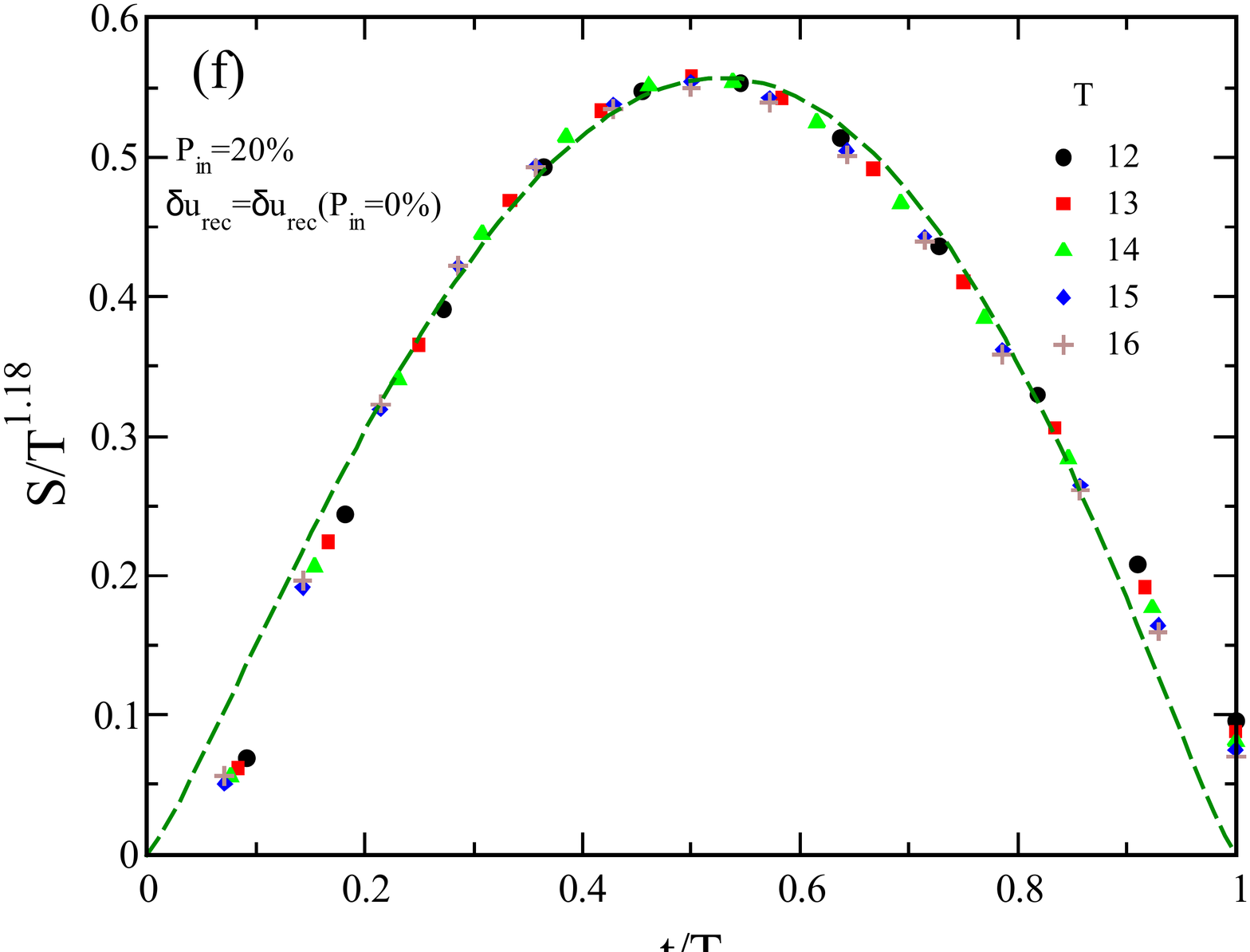}	
	\caption{(a) Average avalanche size $S$ vs. time for different durations in the critical state for $P_{in}=0$. The behaviour of the plots for other inhibitory percentages are similar (not shown here). (b-e) The quantity $S/T^{\gamma_{coll}-1 }$ versus $t/T$ for different $P_{in}$. The average avalanche shape for different durations collapse onto a universal profile. The green dashed line is a parabolic fit to the data. (f) The quantity $S/T^{\gamma_{coll}-1 }$ versus $t/T$ for fixed $\delta u_{rec}=0.001$ and $P_{in}=20\%$. The dashed green line shows the fit with Eq.(\ref{fit}).}
	\label{shape_collapse}
\end{figure*}

As for the self-organized critical model \cite{prl06,chaos}, implementing short term plasticity but with no tuning parameter, the avalanche size distribution exhibits (Fig.\ref{data_collapse}) the scaling behaviour
\begin{equation}
 P(S)\propto S^{-\alpha}f(S(P_{in}\delta u_{rec}(P_{in})/\delta u_{rec}(P_{in}=0))^{\theta})
 \label{scaling_relation}
\end{equation}
where $\delta u_{rec}(P_{in})$ and $\delta u_{rec}(P_{in}=0)$ are the parameter values for the system with a fraction $P_{in}$ of inhibitory neurons 
and for a fully excitatory system to be in the critical state, respectively.
The argument of the scaling function $f(x)$, $x=S(P_{in}\delta u_{rec}(P_{in})/\delta u_{rec}(P_{in}=0))^{\theta}$, now takes into account that the activity is driven in the subcritical regime not only by increasing the fraction of inhibitory neurons but also by not tuning $\delta u_{rec}$: 
The ratio
$\frac{\delta u_{rec}(P_{in})}{\delta u_{rec}(P_{in}=0)}$ is a measure of how far the system is from the critical state. Interestingly, the values of the scaling exponents are in good agreement  with those found for the self-organized model, namely $\alpha\simeq 1.50\pm0.01$ and $\theta\simeq 2.4\pm0.1$. This result suggests that the extension of the scaling regime depends solely on the distance from the critical state and not on the different mechanisms implemented in the neuronal dynamics. 


Within the context of the crackling noise, one has the scaling relation $<S>\sim T^{\gamma}$, with $\gamma$ given in Eq.(\ref{sethna}).
A simple argument to justify this relation is the following \cite{deca}: Consider an avalanche with size $S'$ and duration $T'$ in the scaling regime of the distributions. The probability to observe an avalanche smaller than $S'$ is
$P(S<S')\sim S'^{1-\alpha}$, analogously $P(T<T')\sim T'^{1-\tau}$. If $S'$
is the average size of an avalanche of duration $T'$ and for a narrow distribution of avalanche sizes with fixed duration, then $P(S<S')\approx P(T<T')$
and $T'^{\gamma(1-\alpha)}\sim T'^{1-\tau}$, leading to Eq.(\ref{sethna}).
This relation holds for avalanche sizes and durations in the critical scaling regime and is very robust. It has been extended to a variety of other avalanche processes \cite{alava,brown} and verified experimentally in neuronal systems in vitro and in vivo \cite{beggs12,miller,ponce}. Numerically, Eq.(\ref{sethna}) has been recently validated in networks of integrate and fire neurons \cite{deca} and for the Wilson-Cowan model with different populations of excitatory and inhibitory neurons \cite{plos}.

In Fig.\ref{avgS_T} we show the scaling of the average size with fixed duration vs. $T$ for avalanches in the power-law regime for systems with different percentage of inhibitory neurons. The separate fit of the different curves shows that the slope for all systems is very close to 2, the value expected on the basis of Eq.(\ref{sethna})
for $\alpha=1.5$ and $\tau=2.0$ (see Table 1). Interestingly, this scaling is also verified for systems off-criticality, except for fractions of inhibitory neurons larger than 20\%. We observe in this case that avalanches with a long duration are not observed, which strongly reduces the scaling regime (inset Fig.\ref{avgS_T}). Interestingly, the scaling $<S>\sim T^2$ is still observed but there is no agrement with the Sethna relation Eq.(\ref{sethna}).

\begin{table*}	
	\begin{center}
		\begin{tabular}{||c |c | c| c | c| c|c|c||} 
			\hline
			$P_{in}$ & $\delta u_{rec}$ & $\alpha$ & $\tau$ & $\gamma=\frac{\tau-1}{\alpha-1}$ & $\gamma_{coll}$ &$\gamma_{<S>}$
			&$\gamma_{S(f)}$\\
			
			\hline\hline
			
	0\% & 0.0010 & $1.50\pm0.05$ & $2.05\pm0.05$ &2.1$\pm$0.2  &2.08$\pm$0.02 &$2.10\pm0.05$&$1.98\pm0.05$\\
			\hline
   10\% & 0.0014 & $1.50\pm0.05$ & $2.07\pm0.05$ &2.1$\pm$0.2  &2.14$\pm$0.01 &$2.14\pm0.05$&$1.95\pm0.05$\\
			\hline
   20\% & 0.0023 & $1.50\pm0.05$ & $2.09\pm0.05$ &2.2$\pm$0.2  &2.16$\pm$0.02 &$2.3\pm0.2$&$1.95\pm0.05$\\
			\hline
   30\% & 0.0037 & $1.50\pm0.05$ & $2.15\pm0.05$ &2.3$\pm$0.3  &2.32$\pm$0.03 &$2.4\pm0.2$&$2.00\pm0.05$ \\
			\hline
   10\% & 0.0010 & $1.50\pm0.05$ & $2.1\pm0.1$ &2.2$\pm$0.3  & 2.11$\pm$0.02 &$2.15\pm0.1$&$1.95\pm0.05$\\
			\hline
   20\% & 0.0010 & $1.5\pm0.1$ & $1.9\pm0.2$ &1.8$\pm$0.5  & 2.18$\pm$0.04 & $2.15\pm0.1$&$1.80\pm0.05$\\
			\hline
	30\% &0.0010 & $1.7\pm0.2$ & $1.7\pm0.2$ & 1.0$\pm$0.5 &2.08$\pm$0.04 &$2.1\pm0.1$&$1.3\pm0.1$\\
			\hline
		\end{tabular}
	\end{center}
\caption{Avalanche exponents for different $P_{in}$. The exponents $\tau$ and $\alpha$ have been obtained by fitting the distributions with a power law in the critical case and by a power law truncated by an exponential in the subcritical case. The error on $\gamma$ is obtained by propagation of uncertainty. The exponent of the spectrum $S(f)$ is obtained by fitting with a power law in the high frequency regime. The error on $\gamma_{coll}$ is obtained by exploring the range of exponent values providing good collapse. }
	\label{table1}
\end{table*}

\section{Shape of avalanche profile}
Next, we analyse the shape of the avalanche profiles for fixed avalanche duration in the critical state for different percentage of inhibitory neurons. 
To this end, we consider avalanches, whose sizes are in the scaling regime and have a duration in the range $T\pm 3$, and we plot the average number of firing neurons during the avalanche as function of the rescaled time $t/T$. At criticality, the avalanche profiles are nearly parabolic for all durations (Fig.\ref{shape_collapse}a). According to the scaling of the average
avalanche size with duration, it is possible to rescale the axes to collapse the
curves for different $T$ onto a universal profile, according to $<S(t,T)>=T^{\gamma -1} f(t/T)$. The best collapse would then provide an independent evaluation of the exponent $\gamma$ and therefore a validation of the avalanche  exponents at criticality.
In Figs.\ref{shape_collapse}b-f we show the collapse of  avalanche profiles with different durations for systems with various percentages of inhibitory neurons. The collapse is obtained in all cases with an exponent $\gamma_{coll}\simeq 2$ (see Table 1) and the shape of the profile is well fitted by a parabolic function for all values of $P_{in}$. Interestingly, the same analysis performed off-criticality by fixing $\delta u_{rec}=0.001$, although confirming the value $\gamma_{coll}\simeq 2$, exhibits a profile not compatible with a parabolic function. We have then fitted the profiles with the asymmetric function \cite{alava}
\begin{equation}
 S/T^{\gamma-1}\approx(t/T(1-t/T))^{\gamma-1}(1-a(t/T-0.5)),
 \label{fit}
\end{equation}
where $a$ is the measure of asymmetry. The best collapse (Fig.\ref{shape_collapse}f)
is obtained for $a\simeq-0.25$ for 20\% inhibitory neurons, with rightward asymmetry increasing for larger $P_{in}$ ($a\simeq-0.5$ for $P_{in}=30\%$). Interestingly, a similar rightward asymmetric avalanche shape is obtained in the Wilson-Cowan model for different percentages of inhibitory neurons if the system is set in the critical state by tuning an appropriate parameter \cite{plos}. 
We have tabulated the exponents obtained from different methods in Table-\ref{table1}.   

\begin{figure}[ht]
	\includegraphics[width=0.5\textwidth]{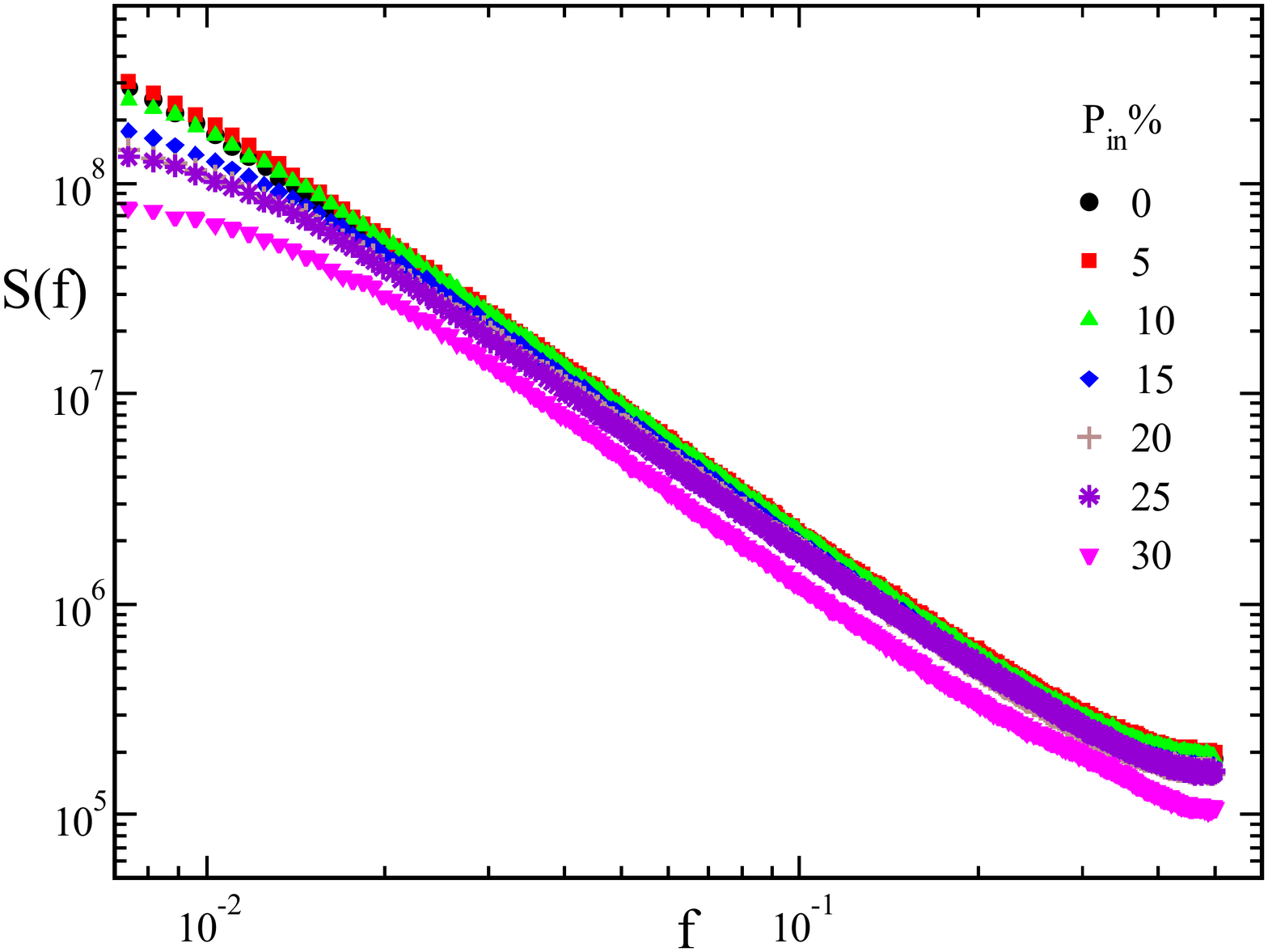}	
	\includegraphics[width=0.5\textwidth]{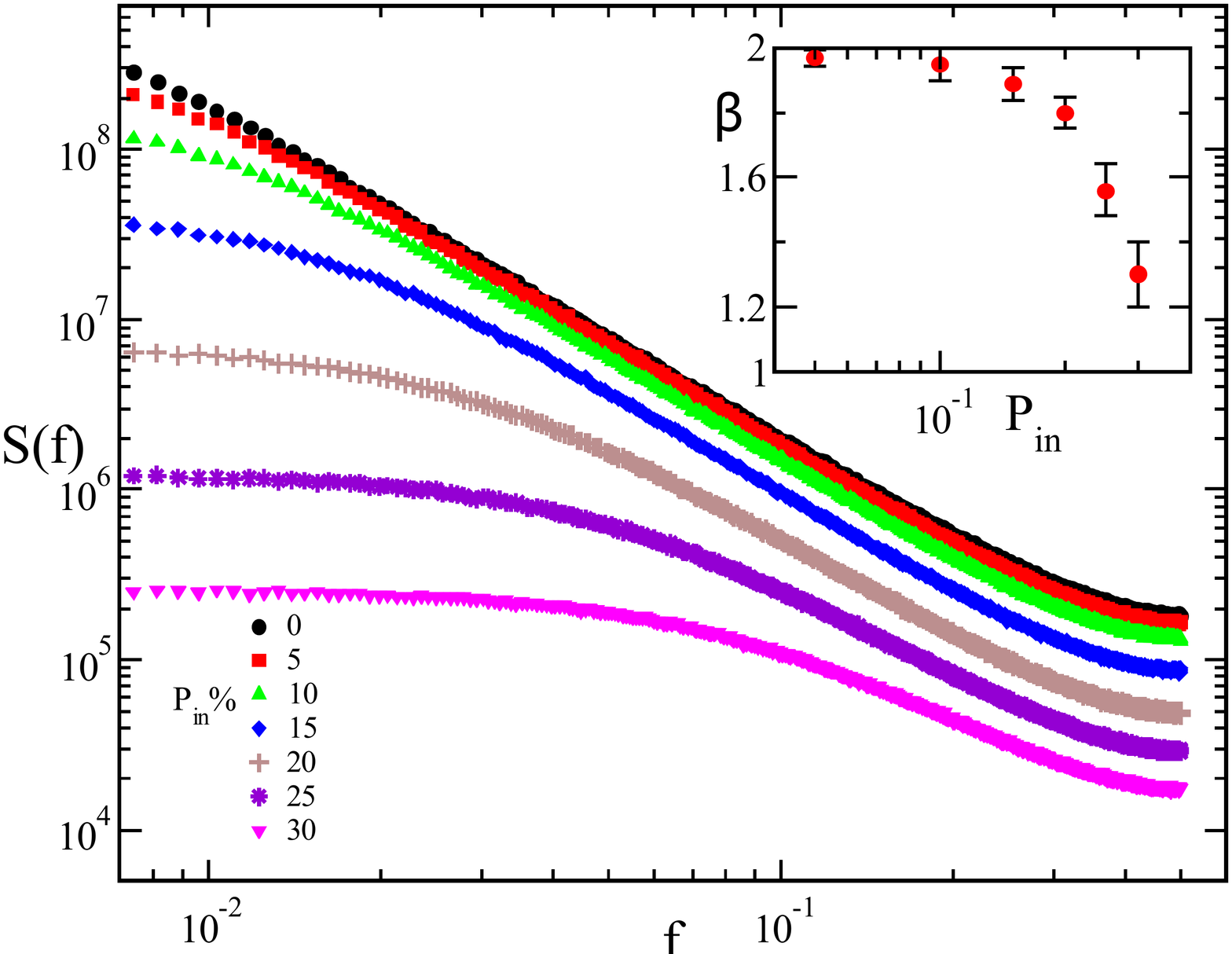}
	\caption{(Top panel) Power spectral density for systems at criticality provides an exponent $\beta\simeq 2.0$ for all fractions of inhibitory neurons. (Bottom panel) Power spectral density for different percentages of inhibitory neurons and fixed $\delta u_{rec}$ corresponding to the critical state for $P_{in}=0\%$. The effective exponent $\beta$ decreases toward unity and the scaling regime shrinks for increasing $P_{in}\to 30\%$. Inset: The effective exponent $\beta$ as function of $P_{in}$.}
	\label{spd}
\end{figure}

\section{Power spectral density}
Next we analyse the behaviour of the activity power spectral density (PSD) in critical and subcritical regimes, searching for the best agreement with experimental data. We consider the temporal sequence of the number of firing neurons $a(t)$ and define the power spectrum as $S(f)=\hat a(f)\hat a^*(f)$, where $\hat a(f)$ is the discrete Fourier transform of $a(t)$ 
\begin{equation}
 \hat a(f)=\sum_{t=0}^{T-1}a(t)e^{-2i\pi f t/T}.
 \label{dft_spd}
\end{equation}
We first analyse the networks at criticality, where $\delta u_{rec}$ is adjusted to the values in Table-\ref{table1} for different $P_{in}$.
Fig.\ref{spd} (top panel) shows the scaling behaviour $S(f)\simeq f^{-\beta}$ which
exhibits a stable critical exponent $\beta\simeq 2.0$ for all percentages of inhibitory neurons, with a crossover towards white noise at small frequencies. This result is fully in agreement with the scaling expected for crackling noise \cite{sethna2001,kuntz}, where the theoretical prediction is 
$S(f)\simeq f^{-\gamma}$ for avalanches at criticality. Unfortunately, this results
does not comply with experimental evidence, which rather exhibits an exponent 
$\beta$ close to unity for a variety of healthy neuronal systems.


Next we evaluate the PSD for the activity signal in networks not tuned at criticality but fixing the parameter $\delta u_{rec}$ at the value for the critical state of $P_{in}=0\%$. Results shown in Fig.\ref{spd} (bottom panel) confirm an effective power law behaviour over a scaling regime decreasing with $P_{in}$ and a crossover towards white noise at small frequencies. Moreover, as shown in the inset, the effective exponent $\beta$ continuously varies with $P_{in}$, from $\beta\simeq 2.0$ for purely excitatory systems to $\beta\simeq 1.0$ for 30\% inhibitory neurons, the fraction typically found in mammal brains (see Table 1). Moreover, the crossover to white noise moves to larger frequencies for increasing $P_{in}$, corresponding to the inverse of the largest avalanche duration in the system. 
The behaviour observed in  Fig.\ref{spd} (bottom panel) has been also evidenced in the self-organized model for neuronal activity \cite{chaos}.
 
\begin{figure}
	\includegraphics[angle=-90,width=0.5\textwidth]{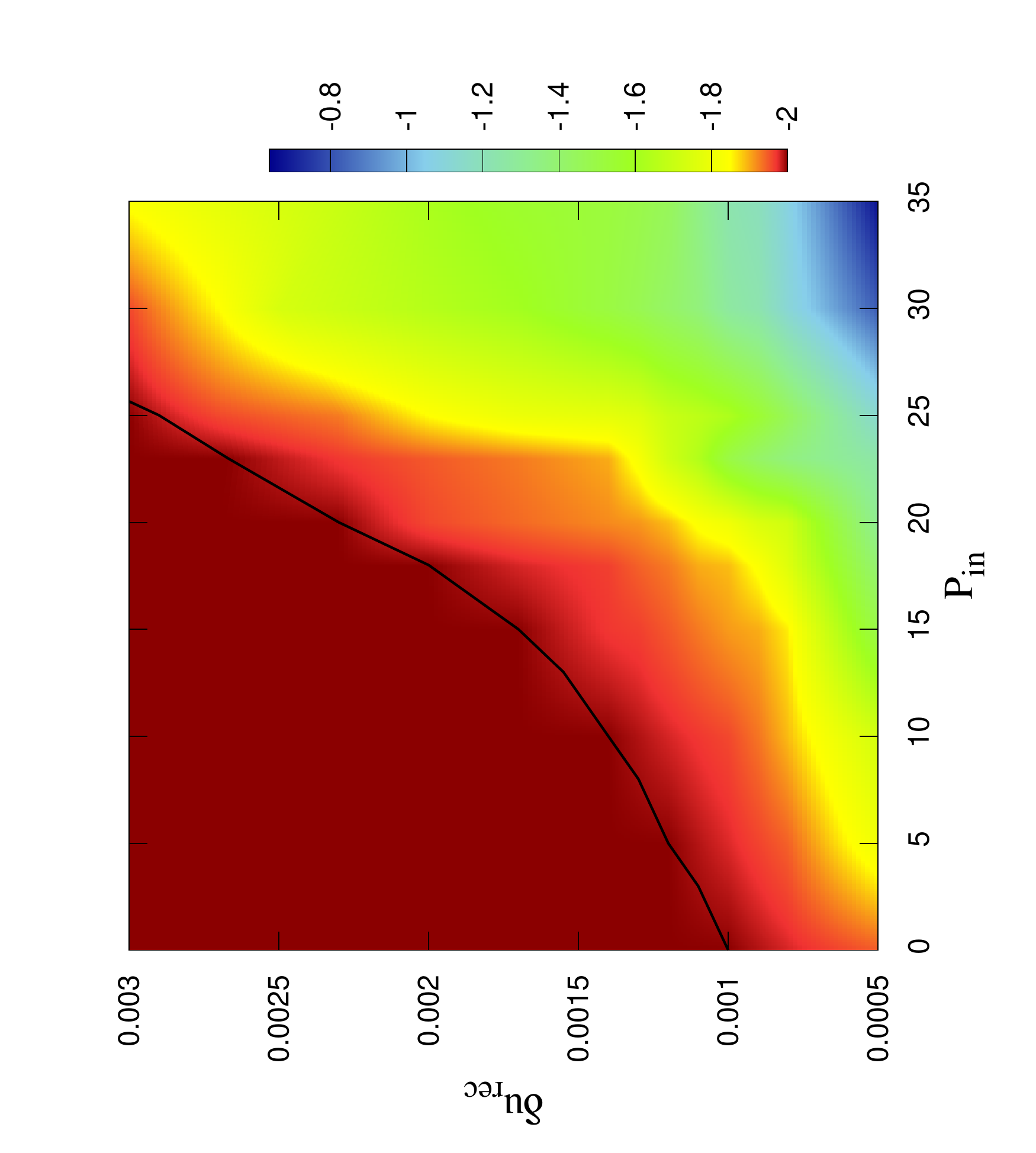}
	\caption{Contour plot of the $\beta$ exponent values for different $\delta u_{rec}$ and $P_{in}$. The value of the exponent is given by the colour code. The continuous black line is the critical line, with the supercritical (subcritical) region for larger (smaller) $\delta u_{rec}$, respectively.
		Irregularities in the contour shape are an artefact due to the numerical discretization.}
	\label{spd_phase}
\end{figure}
To fully characterize the properties of the PSD in the entire phase diagram, we have systematically tuned the parameter $\delta u_{rec}$ over a range of values at and off criticality for different $P_{in}$ and measured the exponent $\beta$. The contour plot in Fig.\ref{spd_phase} suggests an interesting behaviour: At criticality (continuous black line) and in the supercritical phase Brown noise ($\beta\sim2$) is always found, which is in fact the value measured in epileptic patients. Conversely, moving away from the critical line, into the subcritical region, provides $\beta$ values closer to experimental data for healthy patients. The model therefore suggests that systems in healthy conditions act slightly off criticality, in the subcritical regime. 

\section{conclusions}
In neuronal systems inhibition has a crucial role in keeping activity balanced in healthy conditions. Inhibitory neurons act as sinks, hampering activity propagation, therefore their presence affects avalanche dynamics. In particular, the open question is how their dissipative role can affect the predictions by the general scaling theory for avalanches, initially formulated for crackling noise \cite{sethna2001}. Among a variety of scaling relations we investigate the robustness of the predictions on the scaling of the average size with duration, the universal profile and the activity spectra. All these properties, except the last one, have been intensively investigated experimentally and numerically in different contexts, confirming the general validity of the scaling theory. Here we performed a numerical study of a neuronal network model, able to simulate systems at and off criticality and therefore to monitor the role of inhibition. We evidence that the expected scaling behaviour for the average size with duration and the collapse onto a universal profile are robust properties verified in systems with different fractions of inhibitory neurons at and near criticality, provided that we consider avalanches in the scaling regime. Therefore, the arguments of the Sethna scaling appear to be very robust also for neuronal avalanches, confirming the existence of a universal exponent $\gamma\sim 2$ even when the system is not tuned at criticality. In this case the exponents for the size and duration distributions show effective values because of the exponential corrections due to dissipative effects introduced by inhibitory neurons, but these do not affect the scaling of the average size and the avalanche shape, except for the emergence of a slight asymmetry. Interestingly, since the system is off-criticality, the scaling relation Eq.(\ref{sethna}) in this case no longer provides the value of the exponent describing how the average size scales with duration and avalanche profiles collapse.

Moreover, the avalanche scaling theory predicts the same exponent for the power spectrum, $S(f)\sim f^{-\gamma}$ \cite{kuntz}, which is in stark contrast with experimental data for healthy brains, which rather exhibit effectively $1/f$ noise.
We evidence that inhibition is responsible for the crossover in the scaling behaviour of the power spectrum: Without tuning the parameter setting the system at criticality, a fraction of 30\% inhibitory neurons leads to a behaviour closer to $1/f$ noise, same evidence found in self-organized models in absence of a tuning parameter \cite{chaos}. Not compensating inhibition by increasing the parameter $\delta u_{rec}$ amounts to drive the system away from criticality, in the sub-critical regime, as the percentage of inhibitory neurons increases. In this case, effective exponents close to $1/f$ noise can be measured. Therefore, numerical results  
suggest that deviations of $\beta$ from the Brown noise value can provide an indicator of how far from criticality the system operates.

\noindent
\section{Acknowledgements}	LdA would like to thank MIUR project PRIN2017WZFTZP.
AS acknowledges support from MIUR project PRIN201798CZLJ. LdA and AS thank the Program VAnviteLli pEr la RicErca: VALERE 2019 for financial support. HJH thanks FUNCAP and the University of Campania for the visiting professorship. 

\bibliographystyle{h-physrev}
\bibliography{reference_brain}
\end{document}